\documentclass[traditabstract]{aa} 
\usepackage{graphicx,natbib,longtable,txfonts,lscape,lineno,amsmath}

\usepackage[colorlinks=true, pdfstartview=FitV, linkcolor=blue,citecolor=blue, urlcolor=blue]{hyperref}

\newcommand{\FRONE}{FRI{\sl{CAT}}}
\newcommand{\FRTWO}{FRII{\sl{CAT}}}
\newcommand{\WAT}{WAT{\sl{CAT}}}

\usepackage{array}
\newcolumntype{P}[1]{>{\centering\arraybackslash}p{#1}}

\begin{document}
	
\title{Investigating the large-scale environment of wide-angle tailed radio galaxies in the local Universe}

\author{V. Missaglia\inst{1,2,3,4}
	\and 
	A. Paggi\inst{1,3,4,5}
	\and 
	F. Massaro\inst{1,3,4,5}
	\and 
	A. Capetti\inst{3}
	\and 
	R. D. Baldi\inst{6}
	\and
	R. P. Kraft\inst{7}
	\and
	M. Paolillo\inst{8}
	\and
	A. Tramacere\inst{9}
	\and 
	R. Campana\inst{10}
	\and
	I. Pillitteri\inst{11}}

\institute{Dipartimento di Fisica, Universit\`a degli Studi di Torino, via Pietro Giuria 1, 10125, Torino, Italy
	\and
	Institute of Astrophysics, Foundation for Research and Technology - Hellas, Voutes, 7110 Heraklion, Greece
	\and
	INAF - Osservatorio Astrofisico di Torino, via Osservatorio 20, 10025, Pino Torinese, Italy
	\and 
	Istituto Nazionale di Fisica Nucleare, Sezione di Torino, 10125, Torino, Italy
	\and
	Consorzio Interuniversitario per la Fisica Spaziale, via Pietro Giuria 1, 10125, Torino, Italy
	\and
	INAF - Istituto di Radioastronomia, via Piero Gobetti 101, 40129, Bologna, Italy
	\and
	Center for Astrophysics $\mid$ Harvard \& Smithsonian, 60 Garden Street, 02138, Cambridge (MA), USA
	\and
	Dipartimento di Fisica ``Ettore Pancini'', Universit\`a di Napoli Federico II, via Cintia, 80126, Napoli, Italy
	\and
	University of Geneva, Chemin d'Ecogia 16, Versoix, CH-1290, Switzerland
	\and
	INAF/OAS, via Piero Gobetti 101, 40129, Bologna, Italy
	\and
	INAF - Osservatorio Astronomico di Palermo ``G. S. Vaiana'', Piazza del Parlamento 1, 90134, Palermo, Italy
}

\date{\today}

\abstract { 
	We present a statistical analysis of the large-scale (up to 2\,Mpc) environment  of an homogeneous and complete sample, both in radio and optical selection, of wide-angle tailed radio galaxies (WATs) in the local Universe (i.e., with redshifts $z\lesssim$ 0.15). 
	The analysis is carried out using the parameters obtained from cosmological neighbors within 2\,Mpc of the target source. 
	Results on WATs large-scale environments are then compared with that of Fanaroff-Riley type I (FR\,Is) and type II (FR\,IIs) radio galaxies, listed in two others homogeneous and complete catalogs, and selected with the same criterion adopted for the WATs catalog. 
	We obtain indication that at low redshift WATs inhabit environments with a larger number of galaxies than that of FR\,Is and FR\,IIs. 
	In the explored redshift range, the physical size of the galaxy group/cluster in which WATs reside appears to be almost constant with respect to FR\,Is and FR\,IIs, being around 1 Mpc. From the distribution of the concentration parameter, defined as the ratio between the number of cosmological neighbors lying within 500\,kpc and within 1\,Mpc, we conclude that WATs tend to inhabit the central region of the group/cluster in which they reside, in agreement with the general paradigm that WATs are the cluster BCG.
}

	\keywords{surveys -- methods: statistical -- galaxies: active galaxies: clusters: general -- galaxies: jets -- radio continuum: galaxies.}
	\maketitle
	
	\section{Introduction}
	\label{sec:intro} 
	Extragalactic extended radio sources have been classified taking into account the relative positions of their high and low brightness regions, that are found to be correlated with the luminosity of these sources \citep{fanaroff74}: edge-darkened radio sources were classified as Fanaroff-Riley type I (FR\,Is), while edge-brightened as type II (FR\,IIs). A summary of the structural properties of extended extragalactic radio sources is reported by \citet{miley1980}.
	The FR\,I/FR\,II dichotomy reflects the different cosmological evolution of these two classes \citep{longair1971}, and it is debated if there is a link between accretion modes and radio morphology \citep{best2012}. This sources appear also to be related with the large-scale environment where they reside \citep[see e.g.][]{worrall2000}.
Many different methods have been already used to investigate the large-scale environment of radio sources. Thanks to multifrequency observations and redshift estimates, it has been possible to remove unrelated galaxies and so improving the reliability of the analysis \citep{best04}. In the optical range, the Sloan Digital Sky Survey \citep[SDSS,][]{york2000} has expanded our knowledge of galaxy properties, such as luminosities, morphologies, star-formation rates and nuclear activity, and how this properties depend upon the environment that a galaxy inhabits. This can place important constraints on models of galaxy formation and evolution, and allows the intrinsic properties of the galaxies to be separated from those that have been externally induced.
	
Studying the environment of FR\,Is and FR\,IIs on the megaparsec scale, it was found that FR\,Is generally inhabit galaxy-rich environments, being members of groups or galaxy clusters, while FR\,IIs tend to be more isolated as shown, for example, in \citet{zirbel1997}. There are however some well known exceptions, such as Cygnus A \citep{carilli1996}.
	
The environments of powerful radio sources have been widely studied \citep[see e.g.,][]{prestage88,hill91} up to a $z\sim$ 0.5. From the estimate of the galaxy density around these sources, it has been concluded that there is no strong statistical evidence for a difference in the environments hosting FR\,Is and FR\,IIs, but at low-redshift ($z <$ 0.5)  the environments appear less galaxy-rich than that of the counterparts of same radio power at high-redshift.
	
	Another class of radio galaxies are the wide-angle tailed radio galaxies (WATs) that show the so-called ``jet-hotspot-lobe transition'': there are bright hotspots (called  ``warmspots'') closer to their radio core with respect to FR\,IIs and with extended radio plumes beyond them \citep{1990ApJS...72...75O}.
	The general morphology of WATs \citep[firstly classified by][]{owen1976} suggests that these sources interact significantly with the surrounding medium: these sources show bent tails as the result of  the ram pressure due to the relative motion between the radio source and intracluster medium \citep[ICM; see e.g. simulations in][]{massaglia2019}. WATs are normally found in galaxy cluster and are in general associated with the brightest cluster galaxy \citep[BCG; see e.g.][]{burns1981}. This implies that WATs can be found in merging systems, as shown in \citet{gomez1997} via \textit{ROSAT} X-ray data, or relaxed systems showing ``sloshing'' of the central ICM due to cluster minor mergers \citep{asca2006}. WATs have proven to be reliable tracers of high-density environments up to high redshifts (see e.g. \citet{giacintucci2009}) and may be therefore used as probes of the presence of the ICM. In the already cited work by \citet{burns1981}, VLA 20-cm observations of the WAT 1919+479 (4C\,47.51) are presented. The author discuss the morphology, polarization, environment and nature of the galaxy cluster hosting the radio source. The author also highlight that with X-ray observations it is possible to support the contention that the cluster around the WAT is gas rich. 
		
	As shown by \citet{wing2011} through SDSS and Faint Images of the Radio Sky at Twenty Centimeters \citep[FIRST,][]{FIRST} data, bent radio sources are more often found in galaxy clusters than non-bent radio sources, and therefore the authors point out that a radio-selected galaxy cluster sample can be obtained classifying bent radio sources from their radio morphology and then looking for their optical counterpart. In \citet{garon2019}, the authors investigate the effect of the cluster environment on the morphology of the sources, presenting bent sources properties (e.g spatial distribution of radio galaxies around clusters and  bending angle as a function of cluster mass or pressure) selected from the Radio Galaxy Zoo project catalog \citep{zoo2015}. The results  show that the bending is higher in sources near the center of the cluster, but the authors cannot draw firm conclusions on the radio morphology of BCGs hosted in disturbed clusters.

		In the literature, WATs and bent double radio sources environments have been studied both at low ($z\sim$ 0.2) and mid-redshift ($z>$ 0.2). \citet{blanton01} presented observations of a complete, magnitude-limited sample of 40 radio galaxies from the VLA FIRST survey, part of a larger sample of bent-double radio sources at moderate redshift. The most interesting result is that $\sim$46\% of the sources in the sample are associated with groups, some of them being poor groups.  
		The high-redshift Clusters Occupied by Bent Radio AGN (COBRA) Survey \citep{blanton2015} uses bent radio sources as tracers of distant galaxy clusters, on the assumption that, as shown at low-redshift, these sources are good tracers of high-density environments.
		
		\citet{smolcic2007}, having identified a complex galaxy cluster system in the COSMOS field via a WAT, used optical and X-ray data to investigate its host environment. The cluster shows evidence for subclustering, both in diffuse X-ray emission and in the spatial distribution of galaxies found from the optical analysis applying the Voronoi tessellation-based approach.

		In this paper, extending the analysis of the large-scale environment to a complete and homogeneous (both in luminosity and redshift) sample of WATs restricted to the local Universe (i.e., source redshifts $z_{src} \leq$0.15) we can prove if WATs environment differs from that of FR\,Is and FR\,IIs from the catalogs \FRONE\ and \FRTWO. WATs used in this work are listed in the \WAT\ \citep[][details of the catalogs are described in Section~\ref{sec:samples}]{missaglia2019}. A similar analysis is reported in \citet{massaro2019} in which the authors presented the results of the analysis of the large-scale environment of FR\,I and FR\,II radio galaxies from the same catalogs we use here. \citet{massaro2019} concluded that radio galaxies, independently of their radio (FR\,I vs. FR\,II) classification, tend to inhabit galaxy-rich large-scale environments with similar richness. 
		
		Results from previous cited works, even if for samples at higher $z$ with respect to that of the \WAT , show that bent radio morphology is used to identify and characterize the environment in which the bent sources are hosted. For this reason, in this work we have defined parameters that will be used to characterize the environment in which \WAT\ sources lies.
		
		The paper is organized as follows. In \S~\ref{sec:samples} we briefly describe the samples used to carry out our analysis, while in \S~\ref{sec:cn} we provide a brief description of the cosmological neighbors and several ambient parameters obtainable from their distribution. 
		Then \S~\ref{sec:statistics} is devoted to the results of the statistical analysis of the environment of WATs. Finally, summary and conclusions are given in \S~\ref{sec:summary}. In Appendix~\ref{app:figtab} we report values of the parameters obtained from the cosmological neighbors of WATs.
		
		Hereinafter, we adopt cgs units for numerical results and we assume a flat cosmology with $H_0$=69.6 km s$^{-1}$ Mpc$^{-1}$, $\Omega_\mathrm{M}$=0.286 and $\Omega_\mathrm{\Lambda}$=0.714 \citep{bennett14}, unless otherwise stated. Thus, according to these cosmological parameters, in the $z$ range of the \WAT\ , 1\arcsec\ corresponds to 0.408 kpc at $z_\mathrm{src}$=0.02 and to 2.634 kpc at $z_\mathrm{src}$=0.15.

	\section{Sample selection}
	\label{sec:samples} 
	We selected three radio galaxy catalogs to carry out our analysis, all obtained from the radio-loud sample of \citet{best2012}. All sources are selected on the basis of their morphology, as it is shown in the Faint Images of the Radio Sky at Twenty cm (FIRST) radio survey \citep{FIRST}.
	
	The first catalog is the \WAT\ \citep{missaglia2019}, listing 47 radio sources at low redshift ($z_\mathrm{src}\leq$ 0.15) showing two-sided jets with two clear warmspots (i.e., jet knots as bright as 20\% of the nucleus) lying on the opposite side of the radio core and having classical extended emission resembling a plume beyond the warmspots. As shown in \citet{missaglia2019}, WATs show multifrequency properties remarkably similar to FR\,Is radio galaxies, being more powerful at radio wavelengths and similar to FR\,IIs. See Table~\ref{tab:main} for \WAT\ sources radio luminosities at 1.4 GHz.

	The second catalog is the combination of \FRONE\ and s\FRONE\ both described in \citet{capetti2017a}. \FRONE\ sources, chosen on the basis of their FR\,I radio morphology, are selected to have a radio structure extending beyond a distance of 30 kpc from the optical position of the host galaxy. The 14 s\FRONE\ sources have FR\,I radio morphology, whose radio emission extended between 10 and 30 kpc, and are limited to $z_{src}=$0.05 \citep[see][for details]{capetti2017a}. The combination of these two samples includes FR\,Is at redshift $z_{src} \leq 0.15$. Sources from the FR\,I sample have radio luminosities at 1.4 GHz in the range $L_{1.4}\sim 10^{39.5}-10^{41.3}$ erg s$^{-1}$ while sFR\,I radio luminosities spans 10$^{39}-10^{40.4}$ erg s$^{-1}$.
	
	The third catalog is the \FRTWO\ \citep{capetti2017b}, composed of 105 edge-brightened radio sources (FR\,II type) within the same redshift range of the previous catalog.Sources in this sample have radio luminosities at 1.4 GHz that cover the range $L_{1.4}\sim 10^{39.5}-10^{42.5}$ erg s$^{-1}$.

	All three catalogs include only sources lying in the footprint of the SDSS, that is also covered by the main catalog of groups and clusters of galaxies adopted in our analysis: the one created by \citet{tempel2012}, based on a modified version of the Friends-of-Friends algorithm \citep{huchra82,tago10}. The Tempel catalog has the largest number of galaxy cluster/group detections, with spectroscopic redshifts 0.009 $\leq z_\mathrm{cl} \leq $ 0.20, with a peak around 0.08. 
	
	We want to stress that, thanks to the adopted selection criteria, all three radio galaxy catalogs are not contaminated by compact radio objects, as compact steep spectrum sources, and FR\,0s \citep{baldi15,baldi18}, which show a different cosmological evolution and lie in different environments.

	\section{Cosmological neighbors}
	\label{sec:cn} 
	
	To investigate the large-scale environment of WATs we first defined a type of optical sources: the \textit{cosmological neighbors}.
	
	We classify as cosmological neighbors all optical sources lying within a maximum distance of 2\,Mpc radius computed at $z_\mathrm{src}$ of the central radio galaxy, with all the SDSS magnitude flags indicating a galaxy-type object (i.e., {\it uc=rc=gc=ic=zc=}3), and having a spectroscopic redshift $z$ with $\Delta\,z=|z_\mathrm{src}-z|\leq$ 0.005. This value corresponds to $\sim$ 1500 km s$^{-1}$, that is the maximum velocity dispersion in groups and clusters of galaxies \citep[see e.g.,][]{berlind06}. However, in the galaxy-richest clusters, the large velocity dispersions may lead this method to underestimate the local galaxy density, due to some companion galaxies falling outside of the selected range.

	\noindent 
	We indicate as N$_{cn}^{500}$ and N$_{cn}^{2000}$ the number of cosmological neighbors lying within 500\,kpc and 2\,Mpc distance from the central radio galaxy, respectively, that provide an estimate of the environmental richness. 
	
\begin{figure}[!b]
		\begin{center}
			\includegraphics[height=6.6cm,width=8.4cm,angle=0]{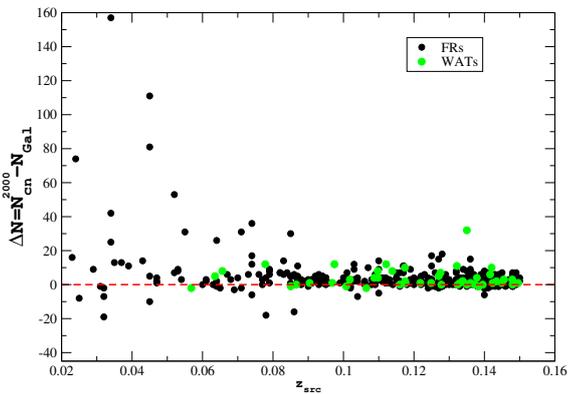}
		\end{center}
		\caption{The difference $\Delta$N between the number of cosmological neighbors lying within 2\,Mpc distance from the central radio source (N$_{cn}^{2000}$), and the richness estimated in the T12 cluster catalog ($N_{gal}$) as function of $z_{src}$. Grey circles mark radio sources in the \FRONE\ , s\FRONE\, and \FRTWO\ , and green diamonds correspond to \WAT\ sources.}
		\label{fig:dN}
	\end{figure}

	As shown in Fig.~\ref{fig:dN}, it is quite evident that the N$_{gal}$ parameter (cluster richness from \citealt{tempel2012}) underestimates the group/cluster richness, and there are only a few cases in which N$_{cn}^{2000}$ provides a lower estimate of the group/cluster richness.

	\section{Parameters definitions}

	\begin{figure}[!b]
		\begin{center}
			\includegraphics[height=6.6cm,width=8.4cm,angle=0]{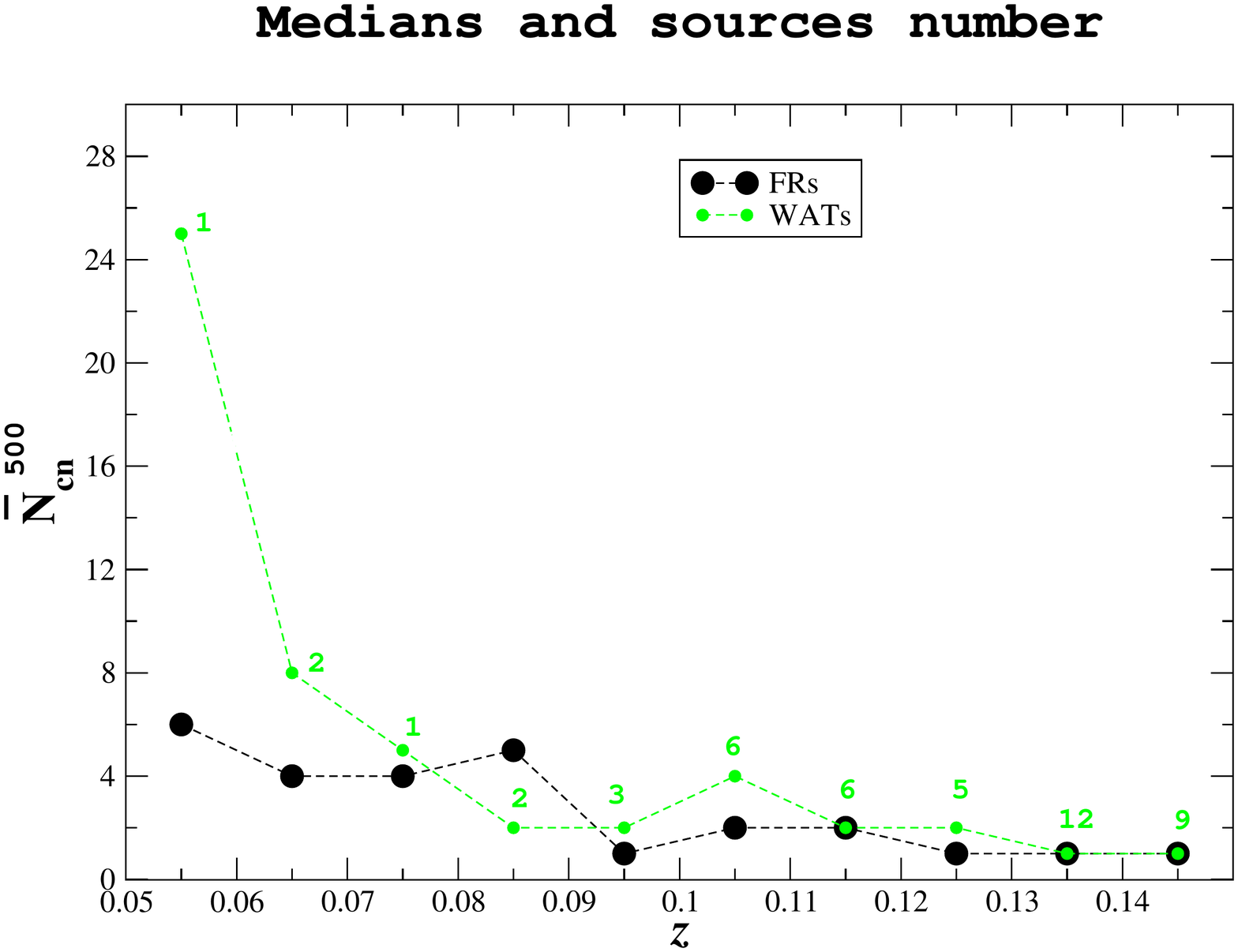}
			\includegraphics[height=6.6cm,width=8.4cm,angle=0]{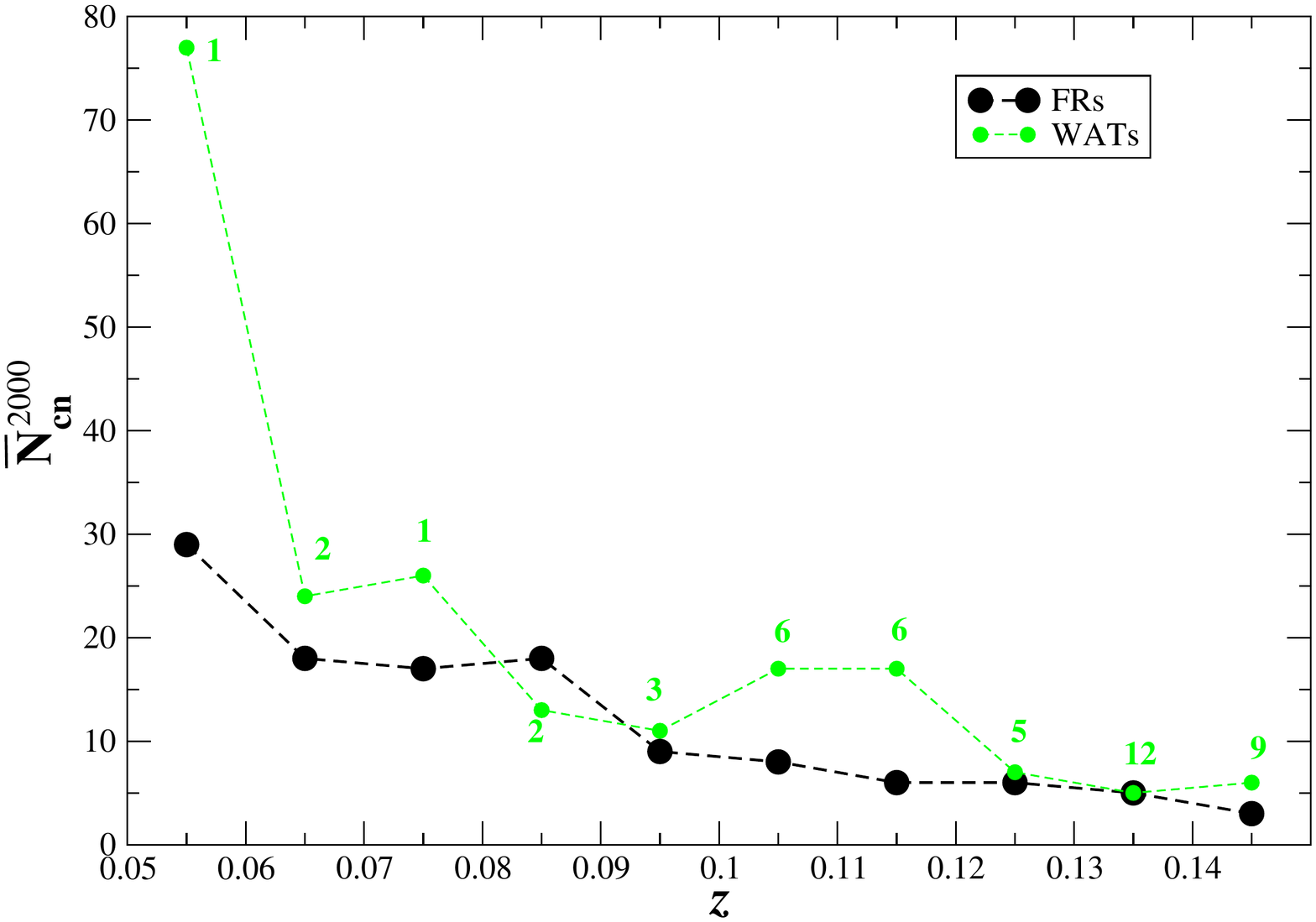}
		\end{center}
		\caption{Median values of the number of cosmological neighbors for the WATs (green) and the FRs (black) within 500\,kpc (upper panel) and 2\,Mpc (lower panel) per bin of
			redshift $z$. Above each median value we also report
			the number of sources in each bin of $z$ starting from 0.05.}
		\label{fig:median}
	\end{figure}

	\label{sec:parameters} 
	Using the distribution of the cosmological neighbors it is possible to define several parameters that can be used to investigate the properties of the large-scale environments of WATs and FR\,Is and FR\,IIs (hereinafter FRs). 
	We thus defined the following quantities:
	\begin{itemize}
		\item The \textit{average projected distance d$_{cn}^{m}$} that is average distance of the distribution of cosmological neighbors within 2\,Mpc from the central radio source. 
		\item The \textit{standard deviation $\sigma_{z}$} of the redshift distribution of the cosmological neighbors surrounding each radio source within 2 Mpc.
		\item The \textit{concentration parameter $\zeta_{cn}$}, defined as the ratio between the number of cosmological neighbors lying within 500\,kpc and within 1\,Mpc. Under the assumption that the cosmological neighbours are uniformly distributed around the radio galaxy analyzed (given that the number of sources around a random position in the sky scales as $N \propto\ \vartheta^2$, where $\vartheta$ is the angular separation from the selected position) we should observe a value of $\zeta_{cn}$ equals 0.25. This parameter allows us to test if the radio galaxy analyzed tends to lie close to the center or in the outskirt of the group or clusters of galaxies in which it resides, if present.
	
	\end{itemize}
	All values of the environmental parameters for the WAT sample described above are reported in Table~\ref{tab:main} in Appendix.

	\section{Statistical analysis}
	\label{sec:statistics} 
	In this section we present the results obtained from the statistical analysis of WATs large scale environment by means of the environmental parameters previously defined, also searching for possible differences between WATs and FRs environmental properties.

	In Fig.~\ref{fig:median} we plot the medians of the number of cosmological neighbors within 500 kpc (upper panel) and 2 Mpc(lower panel) from the \WAT\ sources. 
	We can observe that median values for the WAT sample are systematically higher than that of FRs (with the only exception of the redshift bin 0.08-0.09). If we expect that medians for the WAT sample are distributed randomly, and therefore we have the same probability to find medians values for WATs higher or lower than FRs one, according to the binomial distribution we find that the probability that WATs' N$_{cn}^{500}$ are less than FRs' N$_{cn}^{500}$ is $\sim$ 0.4\% and that WATs' N$_{cn}^{2000}$ are less than FRs' N$_{cn}^{2000}$ is $\sim$ 2\%. 
	Therefore we have some indications that WATs live in galaxy environments richer than that of FRs, but given the small sample of WATs the significance is too low to draw firm conclusions, even if we would expect richer environments, given the general consensus that WATs morphology is due to merger effects.
	
This result is in agreement with the results reported in \citet{golden2021} for high-$z$ bent radio sources, where the authors find that richer clusters host narrower bent radio sources.
	
\begin{figure}[!h]
		\begin{center}
			\includegraphics[height=6.6cm,width=8.4cm,angle=0]{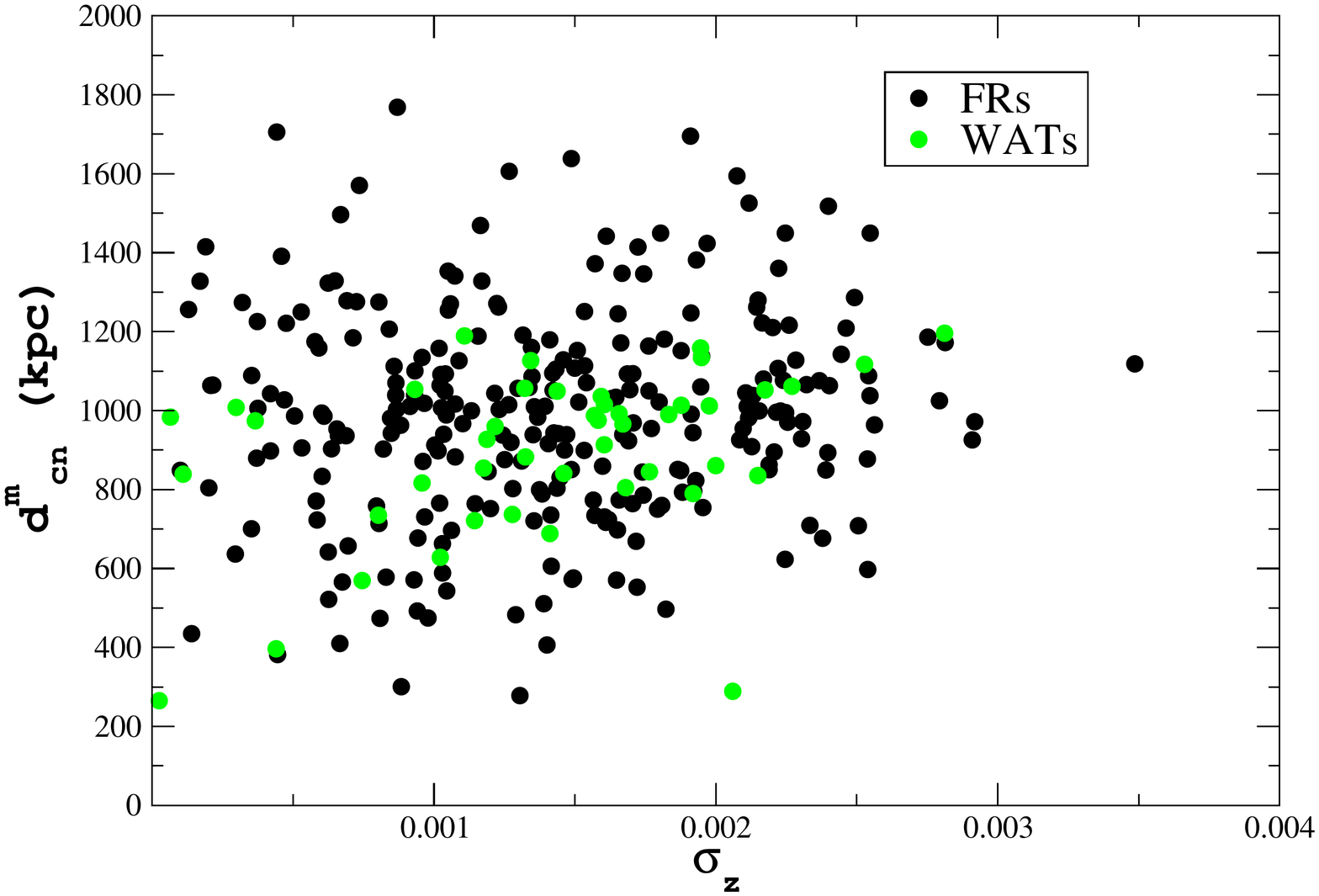}
			\includegraphics[height=6.6cm,width=8.4cm,angle=0]{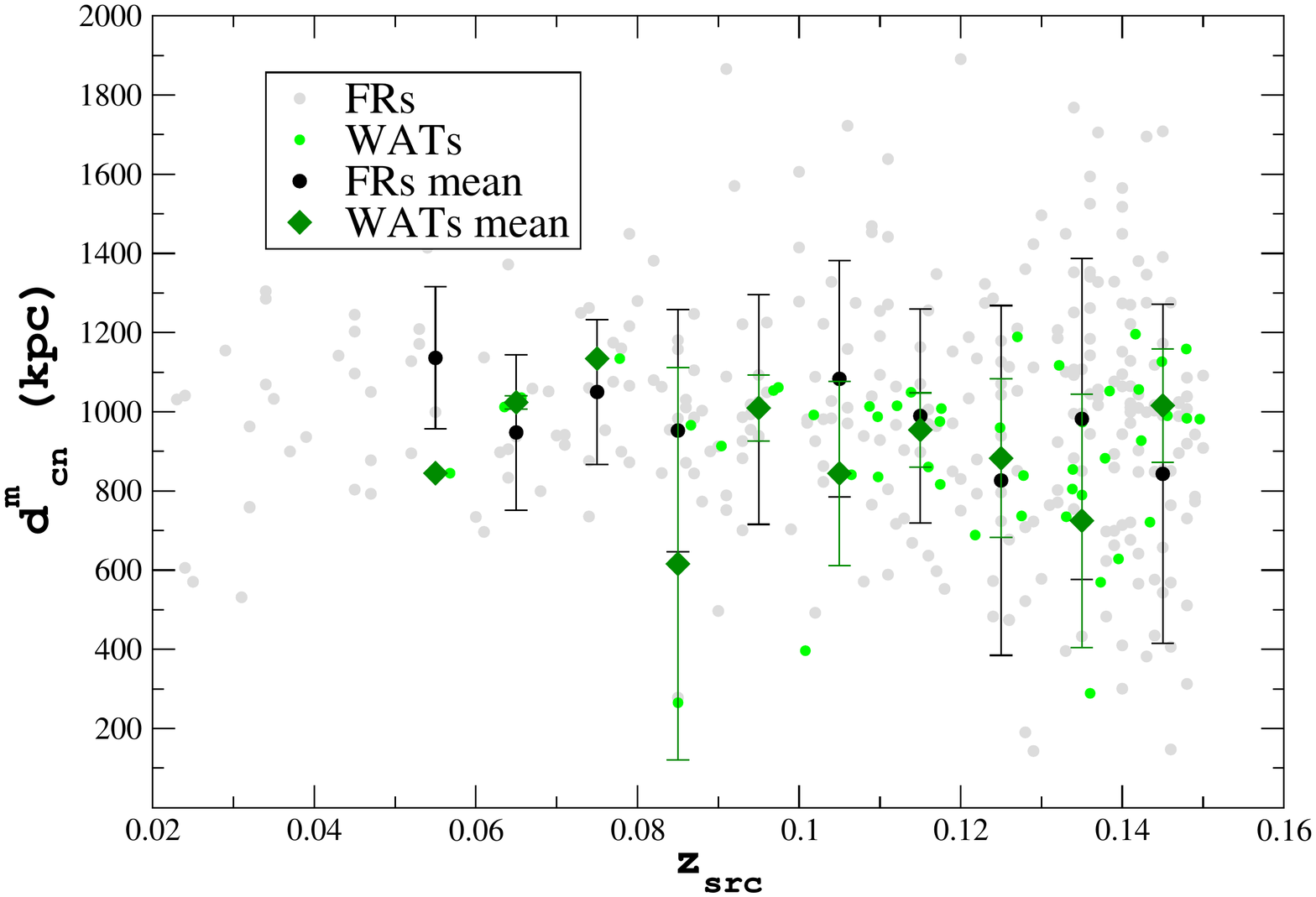}
		\end{center}
		\caption{Average projected distance d$_{cn}^{m}$ of the distribution of cosmological neighbors as function of the standard deviation $\sigma_{z}$ of their redshift distribution (upper panel) and of the redshift $z_{src}$ of the central radio galaxy (lower panel). FR\,Is and FR\,IIs are indicated by black squares on the left and by gray squares on the right. \WAT\ sources are in green. In the left panel we also show the results as mean values per redshift bin. WATs cosmological neighbors are clustered around  d$_{cn}^{m}\sim$ 1 Mpc. In the redshift bin 0.08-0.09 we observe a large scatter because we have only two sources with a large difference in the projected distance of the cosmological neighbors.}
		\label{fig:aver}
	\end{figure}

\begin{figure}[!h]
		\begin{center}
			\includegraphics[height=6.6cm,width=8.4cm,angle=0]{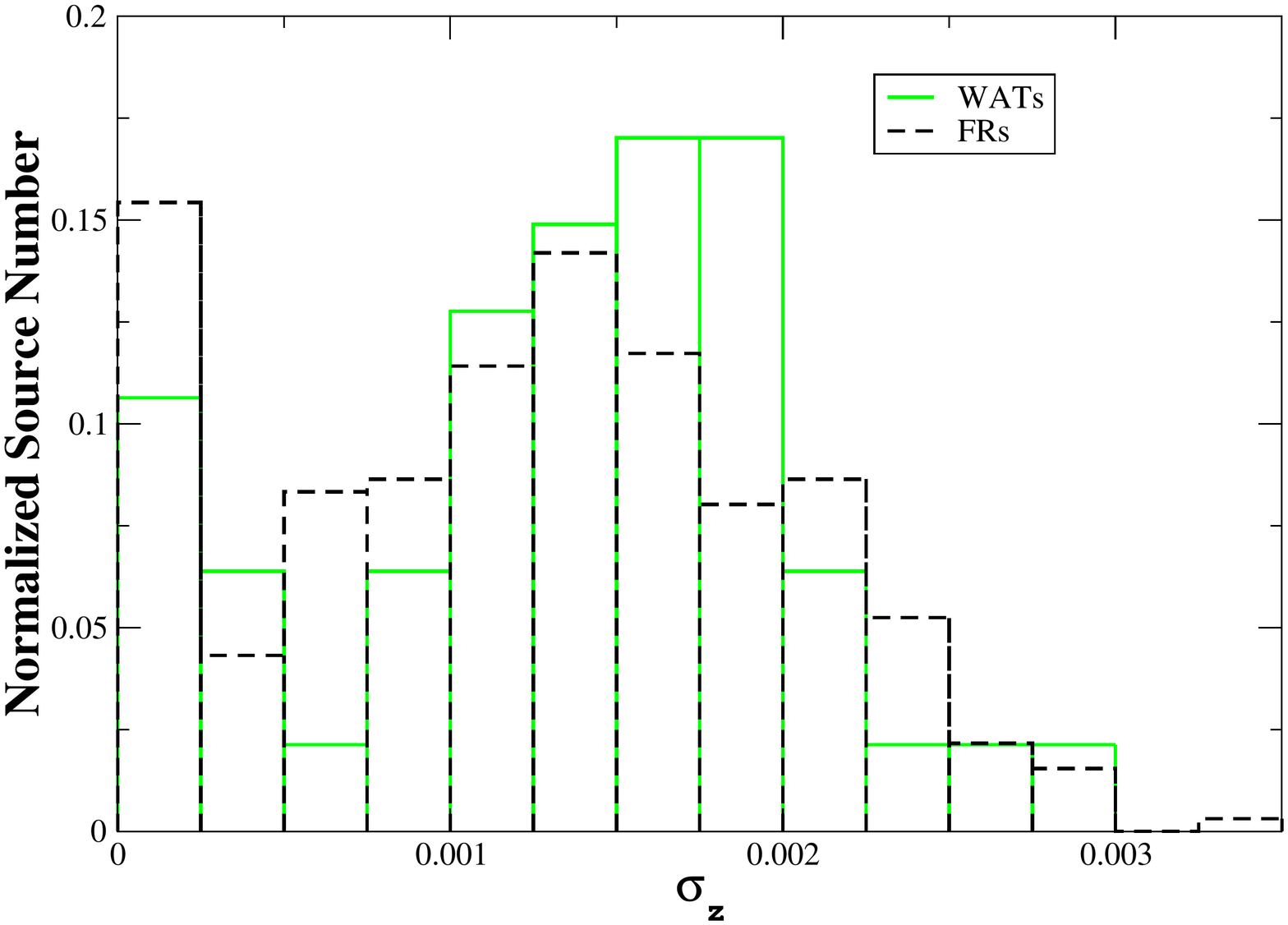}
			\includegraphics[height=6.6cm,width=8.4cm,angle=0]{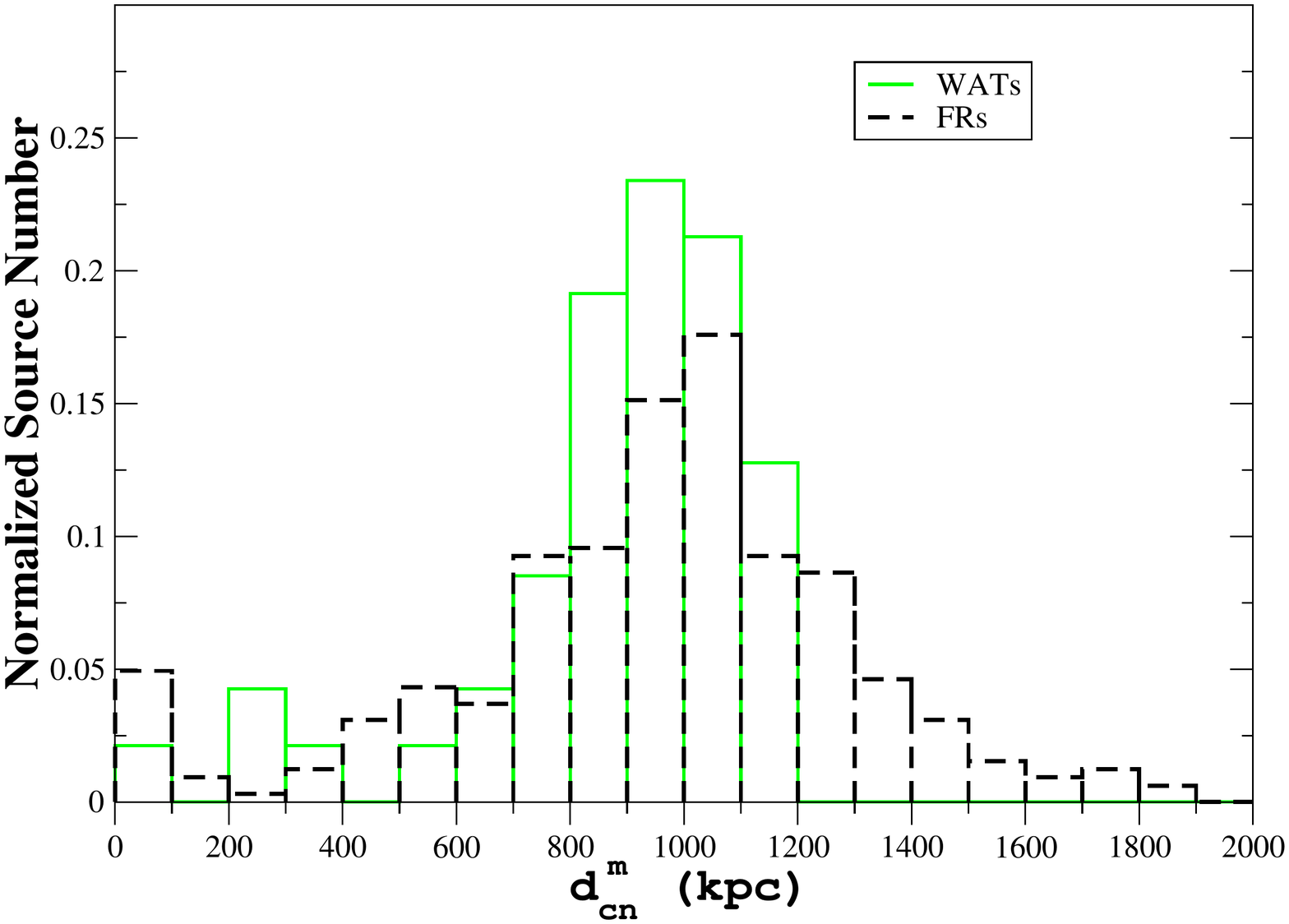}
		\end{center}
		\caption{Distribution of the standard deviation $\sigma_{z}$ of their redshift distribution (upper panel) and of the average projected distance d$_{cn}^{m}$ of the distribution of cosmological neighbors (lower panel). }
		\label{fig:averhist}
	\end{figure}

	Then we explore the distribution of the average projected distance d$_{cn}^{m}$ of cosmological neighbors, which provides an estimate of the galaxy group/cluster physical size, as a function of the standard deviation $\sigma_{z}$ of their redshifts (see Fig.~\ref{fig:aver} upper panel) and as a function of the redshift of the central source $z_{src}$, (see Fig.~\ref{fig:aver} lower panel) where we also overplot the mean value of the projected distance in each redshift bin of 0.01 . In the left panel Fig.~\ref{fig:aver}, we find that the standard deviation $\sigma_{z}$ of cosmological neighbors' redshifts do not exceed a value of 0.003, that is consistent with the threshold of  $\Delta\,z=|z_\mathrm{src}-z|\leq$ 0.005 used to select the cosmological neighbors, while in the right panel we find that all the sources in the \WAT\ are clustered around a value of  d$_{cn}^{m} \sim$ 1 Mpc, implying that the size of the galaxy cluster/group in which the WAT is hosted is constant at low-$z$, and there are no values of d$_{cn}^{m} $ larger than 1.2 Mpc, implying that WATs tend to occupy the central region of the galaxy group/cluster that inhabit. The observed scatter at high-$z$ is due to the low number of cosmological neighbors detected. 
	The same results are reported as normalized distributions in Fig.~\ref{fig:averhist}.

\begin{figure}[!b]
		\begin{center}
			\includegraphics[height=6.6cm,width=8.4cm,angle=0]{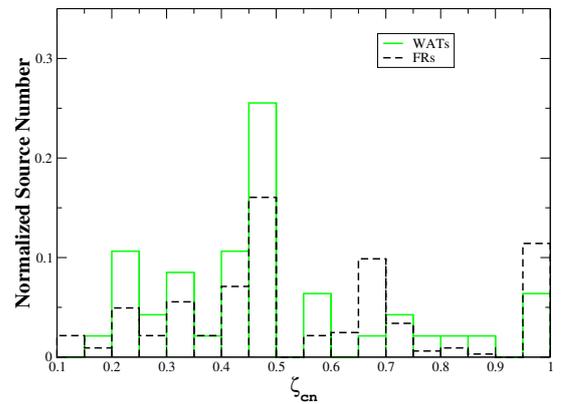}
		\end{center}
		\caption{The distribution of the concentration parameter $\zeta_{cn}$ estimated using the distribution of the cosmological neighbors. Black histogram refers to the FR\,I and FR\,IIs populations, and green one to WATs. The largest fraction of their values lie above the value of 0.25 expected assuming a uniform distribution of cosmological neighbors. }
		\label{fig:ccn}
	\end{figure}

In Fig.~\ref{fig:ccn} we show the distribution of the concentration parameter $\zeta_{cn}$ for \WAT\ and FRs sources. If we assume that the cosmological neighbors are uniformly distributed around the radio galaxy analyzed, we should observe a value of $\zeta_{cn}$= 0.25. We obtain, instead, that the majority of the \WAT\ sources (41 sources out of 47) have a value of $\zeta_{cn}$ higher than 0.25. This means that WATs tend to occupy the central region of the galaxy group/cluster in which they reside, in agreement with WATs being the BCGs. As shown in \citet{golden2021} among the bent high-$z$ radio sources considered in their study, some are indeed BCGs and others may evolve into BCGs.
However, the lack of X-ray observation prevents us to compare the ICM morphology with the spatial distribution of the cosmological neighbors. As pointed out in \citet{vardoulaki2019}, the radio morphology, and in particular how the morphology is disturbed, can be used to identify possible X-ray group previously unidentified.
As shown in \citet{morris2022} bent sources are  usually hosted in groups with an higher galaxy density with respect to that hosting non-bent sources, and in general bent sources are hosted in environments that are larger, denser and less relaxed that unbent sources. As pointed out by the authors, also these results would benefit from X-ray observations, to trace the ICM responsible of the bending.

	\section{Summary and Conclusions}
	\label{sec:summary} 
	In this paper we presented an extensive investigation of WATs large-scale environment in the local Universe (i.e., at $z_{src} \leq$0.15). Our analysis made use of cosmological neighbors, defined as optical sources lying within 2\,Mpc from the target source, and with a redshift difference $\Delta\,z\leq0.005$ with respect to the radio galaxy lying at the center of the field examined. 
	In our study we also compared the large-scale environments of those radio galaxies classified as WATs with that of FR\,Is and FR\,IIs, all selected from extremely homogeneous catalogs, with uniform radio, infrared and optical data available for all sources. For FR\,Is and FR\,IIs it has been found that, independently of their radio morphological classification, they all have environments that are indistinguishable. For this reason, we aimed at investigating the environmental properties of our sample with those already established for FR\,Is and FR\,IIs. We also want to stress that this analysis can not provide information on the intrinsic differences of the cluster hosting FRs and WAT, if any.
	
	We emphasize the importance of {\it comparing radio sources in the same redshift bins} to obtain a complete overview of their large-scale environments, because this method takes into account cosmological biases.

	Our main results are summarized as follows:
	
	\begin{enumerate}
		\item Median values of the number of cosmological neighbors within 500 kpc e 2 Mpc (N$_{cn}^{500}$ and N$_{cn}^{2000}$) are systematically higher than those of radio galaxies within a level of confidence of 0.4\% and 2\%, depending on N$_{cn}^{500}$ and N$_{cn}^{2000}$, respectively;
		\item The average projected distance d$_{cn}^{m}$ of the cosmological neighbors as function of the standard deviation $\sigma_{z}$ of the redshift distribution of the cosmological neighbors and of the redshift $z_{src}$ of the sources is clustered around a distance of $\sim$1 Mpc, impling  that in the redshift range explored, WATs environments have similar sizes, and do not exceed 1.2 Mpc, while there is no trend observed for FR\,Is and FR\,IIs;
		\item Typical values of the concentration parameter $\zeta_{cn}$ for WATs are well above 0.25 (it is $\geqslant$ 0.25 for 41 sources out of 47), value expected considering a uniform distribution of cosmological neighbors around the central RG, implying that WATs tend to inhabit the central region of the galaxy group/cluster in which they reside, therefore possibly being associated with the BCG of the galaxy group/cluster.
	
	\end{enumerate}
	
	We plan to extend our sample with observations from low radio frequency telescopes, such as LOFAR and the uGMRT, augmented with the analysis of \WAT\ X-ray observations, as the ones that eROSITA could provide in the upcoming future. We could therefore estimate properties of the ICM, such as X-ray luminosity $L_X$, mass and environmental mass $M_{env}$ as well as X-ray fluxes. This information will complement the results obtained from the environmental parameters, given that both d$_{cn}^{m}$ and $\zeta_{cn}$ traces the position of the mass, not the gas. 
We also want to highlight that we have explored any link between $d_m^{cn}$ and $\sigma_{z}$ in comparison with the absolute magnitude in the $R$ band of the radio galaxy, but no trend/link is identified. A similar situation occurs also when comparing both these environmental parameters with radio power $L_R$ and emission line luminosity of the [OIII], i.e., L$_{\rm[OIII]}$. These results can be compared with that presented in \citet{croston2019} for a sample of radio-loud AGN in the LOFAR Two-Metre Sky Survey (LoTSS) Data Release 1 catalogues. The authors find trends between the richness of the cluster and the radio luminosity, and also investigated the position of the sources with respect to the cluster centre. The absence of a trend in our sample could provide new insights on the different density of the environment observed for FRs and WATs.

	\appendix
	\section{Table}
	\label{app:figtab}

	We report here in Table~\ref{tab:main} all parameters used for the analysis of the WATs presented here.

	\begin{table*}[]
		\caption{Properties of the \WAT\ sources presented in this work with environmental parameters values.}
	\begin{center}
		\begin{tabular}{|l|r|r|r|r|r|r|r|}
			\hline
			\multicolumn{1}{|c|}{SDSS name} &
			\multicolumn{1}{c|}{$z_{src}$} &
			\multicolumn{1}{c|}{log $\nu L_{\nu}$} &
			\multicolumn{1}{c|}{N$_{cn}^{500}$ } &
			\multicolumn{1}{c|}{N$_{cn}^{2000}$ } &
			\multicolumn{1}{c|}{$\zeta_{cn}$} &
			\multicolumn{1}{c|}{$\sigma_z$}  &
			\multicolumn{1}{c|}{$d_{cn}^{m}$ } \\
			\hline
			
			J004312.85-103956.0 & 0.12754 &  40.93  & 2.0  & 7.0   & 0.40 & 0.001 & 736.99 \\
			J080101.35+134952.2 & 0.10872 & 41.21 & 5.0  & 16.0  & 0.55  & 0.002 & 1013.38 \\
			J080337.67+105042.4 & 0.14234 & 40.98 & 1.0 & 4.0  & 0.33 & 0.001 & 927.21  \\
			J081803.86+543708.4 & 0.11742 & 41.43 & 7.0 & 28.0  & 0.5  & 0.002 & 975.40 \\
			J082718.31+463510.8 & 0.12487 & 40.74 & 2.0  & 9.0   & 0.33 & 0.001 & 959.80 \\
			J085116.24+082723.1 & 0.06354 & 40.36 & 7.0  & 15.0  & 0.87 & 0.002 & 1011.96 \\
			J091337.21+031720.5 & 0.14163 & 40.94 &  1.0 & 10.0  & 0.25 & 0.003 & 1195.88 \\
			J092428.89+141409.3 & 0.13842 & 40.44 & 5.0 & 33.0  & 0.42 & 0.002 & 1052.34 \\
			J092539.06+362705.6 & 0.11212 & 41.49 & 3.0  & 18.0  & 0.37 & 0.002 & 1015.23 \\
			J092612.34+324721.2 & 0.13953 & 41.38 & 3.0  &  4.0   & 1.0 & 0.001 & 628.37 \\
			J093349.82+451957.8 & 0.13387 & 41.33 & 1.0 & 4.0   & 0.5  & 0.001 & 854.25 \\
			J095716.41+190651.2 & 0.09039 & 40.88 & 2.0 & 11.0  & 0.29 & 0.002 & 913.53 \\
			J101932.33+140301.8 & 0.14558 & 41.33 & 4.0 & 10.0  & 0.80 & 0.002 & 989.99 \\
			J103502.62+425548.3 & 0.13602 & 40.86 & 2.0  & 3.0   & 0.67 & 0.002 & 288.96\\
			J103605.76+000606.8 & 0.09683 & 41.25 & 0.0  & 3.0 & 0.0 & 9.32E-4 & 1053.76 \\
			J103636.24+383508.1 & 0.14489 & 41.14 & 0.0 & 3.0  & 0.0 & 0.001 & 1126.64 \\
			J103856.37+575247.5 & 0.10078 & 40.95 & 1.0  & 2.0 & 0.5 & 4.40E-4 & 396.52 \\
			J104645.86+314426.8 & 0.11386 & 40.81 & 2.0 & 16.0 & 0.28 & 0.001 & 1049.01 \\
			J104914.08+005945.2 & 0.10648 & 40.63 & 9.0 & 22.0  & 0.56 & 0.001 & 840.99 \\
			J114020.23+535029.1 & 0.14799 & 41.04 &0.0 & 2.0 & 0.0 & 6.50E-5 & 983.40 \\
			J114111.81+054405.0 & 0.09743 & 41.24 & 13.0 & 57.0 & 0.59 & 0.002 & 1061.37 \\
			J115424.56+020653.0 & 0.13243 & 40.64 & 0.0  & 0.0   & 0.0 &  0.0 & 0.0  \\
			J115513.65-003133.9 & 0.13218 & 41.02 & 2.0 & 14.0  & 0.33  & 0.002 & 1117.27 \\
			J120118.19+061859.3 & 0.13505 & 40.54 & 1.0  & 3.0  & 0.5 & 3.66E-4  & 974.12 \\
			J120455.02+483256.9 & 0.06562 & 40.62 & 10.0 & 32.0  & 0.77  & 0.002 & 1035.64 \\
			J121439.53+052803.9 & 0.0778  & 40.13 & 5.0  & 26.0  & 0.5  & 0.002 & 1134.36 \\
			J130904.46+102935.3 & 0.08661 & 41.17 & 3.0  & 23.0  & 0.19 & 0.002 & 965.99 \\
			J133038.38+381609.7 & 0.10978 & 40.86 & 3.0 & 7.0  & 0.75 & 0.002 & 835.37 \\
			J135315.36+550648.2 & 0.14342 & 40.66 & 3.0 & 6.0 & 0.75 & 0.001 & 721.07 \\
			J141456.58+001223.0 & 0.12702 & 40.82 & 1.0  &  9.0  & 0.25 & 0.001 & 1189.28 \\
			J141513.98-013703.7 & 0.14959 & 41.24 & 0.0	 & 1.0  & 0.0 & 0.0 & 981.44 \\
			J141718.94+060812.3 & 0.10972 & 40.91 & 3.0  & 17.0  & 0.30 & 0.002 & 987.50 \\
			J141731.27+081230.1 & 0.0568  & 40.58 & 25.0 & 77.0  & 0.48 & 0.002 & 844.98 \\
			J141927.23+233810.2 & 0.13732 & 40.12 & 1.0 & 2.0 & 0.5 & 7.45E-4 & 569.33  \\
			J143304.34+033037.6 & 0.14792 & 40.88 & 1.0 & 6.0  & 0.5 & 0.002 & 1158.48\\
			J143409.03+013700.9 & 0.13786 & 41.61 & 2.0 & 9.0 & 0.40 & 0.001 & 882.68 \\
			J144700.45+460243.5 & 0.12777 & 41.04 & 1.0 & 2.0  & 1.0 & 1.10E-4 & 838.92 \\
			J144904.27+025802.7 & 0.12179 & 40.36 & 2.0  & 7.0  & 0.40  & 0.001 & 688.48 \\
			J150229.04+524402.0 & 0.13307 & 41.10 & 1.0 & 5.0  & 0.25  & 8.03E-4 & 734.93 \\
			J151108.77+180153.3 & 0.116 & 41.28 & 6.0 & 19.0  & 0.5  & 0.002 & 860.58 \\
			J154346.14+341521.6 & 0.11747 & 40.20 & 2.0 & 7.0  & 0.5 & 9.58E-4 & 816.44  \\
			J154729.59+145657.0 & 0.08501 & 40.69 & 2.0 & 2.0  & 1.0  & 2.50E-5 & 265.28 \\
			J155343.59+234825.4 & 0.11761 & 41.28 & 1.0 & 3.0 & 0.5 & 2.98E-4 & 1008.10 \\
			J161828.98+295859.6 & 0.13382 & 40.96 &  1.0  & 5.0 & 0.25 & 0.002 & 804.95 \\
			J164527.68+272005.8 & 0.10182 & 40.65 & 5.0 & 20.0  & 0.45 & 0.002 & 991.86 \\
			J212546.35+005551.8 & 0.13501 & 41.62 & 10.0 & 32.0 & 0.42 & 0.002 & 789.58 \\
			J222455.24-002302.3 & 0.14204 & 40.64 & 1.0 &  10.0 & 0.25 & 0.001 & 1056.20 \\
			\hline
		\end{tabular}
	\end{center}
	\label{tab:main}

	Col. (1): SDSS name. \\
	Col. (2): source redshift. \\
	Col. (3): logarithm of the radio luminosity (erg s$^{-1}$).\\
	Col. (4,5): Number of cosmological neighbors within 500 and 2000 kpc, respectively, estimated at the $z_{src}$ of the central radio galaxy.\\
	Col. (6): The concentration parameter $\zeta_{cn}$.\\
	Col. (7): The standard deviation of the redshift distribution for the cosmological neighbors within 2\,Mpc.\\
	Col. (8): Average projected distance of cosmological neighbors within 2 Mpc. \\
	
\end{table*}
	
	\newpage
We thank the anonymous referee for useful comments that led to improvements in the paper. 
	F. M. wishes to thank Dr. C. C. Cheung for their valuable discussions on this project initially planned during the IAU 313 on the Galapagos islands. 
	This work is supported by the ``Departments of Excellence 2018 - 2022'' Grant awarded by the Italian Ministry of Education, University and Research (MIUR) (L. 232/2016). This research has made use of resources provided by the Compagnia di San Paolo for the grant awarded on the BLENV project (S1618\_L1\_MASF\_01) and by the Ministry of Education, Universities and Research for the grant MASF\_FFABR\_17\_01. This investigation is supported by the National Aeronautics and Space Administration (NASA) grants GO4-15096X, AR6-17012X and GO6-17081X. F.M. acknowledges financial contribution from the agreement ASI-INAF n.2017-14-H.0.
	Funding for SDSS and SDSS-II has been provided by the Alfred P. Sloan Foundation, the Participating Institutions, the National Science Foundation, the U.S. Department of Energy, the National Aeronautics and Space Administration, the Japanese Monbukagakusho, the Max Planck Society, and the Higher Education Funding Council for England. The SDSS Web Site is http://www.sdss.org/. The SDSS is managed by the Astrophysical Research Consortium for the Participating Institutions. The Participating Institutions are the American Museum of Natural History, Astrophysical Institute Potsdam, University of Basel, University of Cam- bridge, Case Western Reserve University, University of Chicago, Drexel University, Fermilab, the Institute for Advanced Study, the Japan Participation Group, Johns Hopkins University, the Joint Institute for Nuclear Astrophysics, the Kavli Institute for Particle Astrophysics and Cosmology, the Korean Scientist Group, the Chinese Academy of Sciences (LAMOST), Los Alamos National Laboratory, the Max- Planck-Institute for Astronomy (MPIA), the Max-Planck- Institute for Astrophysics (MPA), New Mexico State University, Ohio State University, University of Pittsburgh, University of Portsmouth, Princeton University, the United States Naval Observatory, and the University of Washington. TOPCAT and STILTS astronomical software \citep{taylor05} were used for the preparation and manipulation of the tabular data and the images.

	\bibliographystyle{aa}
	\typeout{}
	\bibliography{biblio.bib}

\begin{thebibliography}{41}
\expandafter\ifx\csname natexlab\endcsname\relax\def\natexlab#1{#1}\fi

\bibitem[{{Ascasibar} \& {Markevitch}(2006)}]{asca2006}
{Ascasibar}, Y. \& {Markevitch}, M. 2006, \apj, 650, 102

\bibitem[{{Baldi} {et~al.}(2015){Baldi}, {Capetti}, \& {Giovannini}}]{baldi15}
{Baldi}, R.~D., {Capetti}, A., \& {Giovannini}, G. 2015, \aap, 576, A38

\bibitem[{{Baldi} {et~al.}(2018){Baldi}, {Capetti}, \& {Massaro}}]{baldi18}
{Baldi}, R.~D., {Capetti}, A., \& {Massaro}, F. 2018, \aap, 609, A1

\bibitem[{Banfield {et~al.}(2015)Banfield, Wong, Willett, Norris, Rudnick,
  Shabala, Simmons, Snyder, Garon, Seymour, Middelberg, Andernach, Lintott,
  Jacob, Kapińska, Mao, Masters, Jarvis, Schawinski, Paget, Simpson,
  Klöckner, Bamford, Burchell, Chow, Cotter, Fortson, Heywood, Jones, Kaviraj,
  López-Sánchez, Maksym, Polsterer, Borden, Hollow, \& Whyte}]{zoo2015}
Banfield, J.~K., Wong, O.~I., Willett, K.~W., {et~al.} 2015, Monthly Notices of
  the Royal Astronomical Society, 453, 2326

\bibitem[{{Becker} {et~al.}(1995){Becker}, {White}, \& {Helfand}}]{FIRST}
{Becker}, R.~H., {White}, R.~L., \& {Helfand}, D.~J. 1995, \apj, 450, 559

\bibitem[{{Bennett} {et~al.}(2014){Bennett}, {Larson}, {Weiland}, \&
  {Hinshaw}}]{bennett14}
{Bennett}, C.~L., {Larson}, D., {Weiland}, J.~L., \& {Hinshaw}, G. 2014, \apj,
  794, 135

\bibitem[{{Berlind} {et~al.}(2006){Berlind}, {Frieman}, {Weinberg}, {Blanton},
  {Warren}, {Abazajian}, {Scranton}, {Hogg}, {Scoccimarro}, {Bahcall},
  {Brinkmann}, {Gott}, {Kleinman}, {Krzesinski}, {Lee}, {Miller}, {Nitta},
  {Schneider}, {Tucker}, {Zehavi}, \& {SDSS Collaboration}}]{berlind06}
{Berlind}, A.~A., {Frieman}, J., {Weinberg}, D.~H., {et~al.} 2006, \apjs, 167,
  1

\bibitem[{{Best}(2004)}]{best04}
{Best}, P.~N. 2004, \mnras, 351, 70

\bibitem[{{Best} \& {Heckman}(2012)}]{best2012}
{Best}, P.~N. \& {Heckman}, T.~M. 2012, \mnras, 421, 1569

\bibitem[{{Blanton} {et~al.}(2001){Blanton}, {Gregg}, {Helfand}, {Becker}, \&
  {Leighly}}]{blanton01}
{Blanton}, E.~L., {Gregg}, M.~D., {Helfand}, D.~J., {Becker}, R.~H., \&
  {Leighly}, K.~M. 2001, \aj, 121, 2915

\bibitem[{{Blanton} {et~al.}(2015){Blanton}, {Paterno-Mahler}, {Wing}, {Ashby},
  {Golden-Marx}, {Brodwin}, {Douglass}, {Randall}, \& {Clarke}}]{blanton2015}
{Blanton}, E.~L., {Paterno-Mahler}, R., {Wing}, J.~D., {et~al.} 2015, in
  Extragalactic Jets from Every Angle, ed. F.~{Massaro}, C.~C. {Cheung},
  E.~{Lopez}, \& A.~{Siemiginowska}, Vol. 313, 315--320

\bibitem[{{Burns}(1981)}]{burns1981}
{Burns}, J.~O. 1981, \mnras, 195, 523

\bibitem[{{Capetti} {et~al.}(2017{\natexlab{a}}){Capetti}, {Massaro}, \&
  {Baldi}}]{capetti2017a}
{Capetti}, A., {Massaro}, F., \& {Baldi}, R.~D. 2017{\natexlab{a}}, \aap, 598,
  A49

\bibitem[{{Capetti} {et~al.}(2017{\natexlab{b}}){Capetti}, {Massaro}, \&
  {Baldi}}]{capetti2017b}
{Capetti}, A., {Massaro}, F., \& {Baldi}, R.~D. 2017{\natexlab{b}}, \aap, 601,
  A81

\bibitem[{{Carilli} \& {Barthel}(1996)}]{carilli1996}
{Carilli}, C.~L. \& {Barthel}, P.~D. 1996, \aapr, 7, 1

\bibitem[{{Croston} {et~al.}(2019){Croston}, {Hardcastle}, {Mingo}, {Best},
  {Sabater}, {Shimwell}, {Williams}, {Duncan}, {R{\"o}ttgering}, {Brienza},
  {G{\"u}rkan}, {Ineson}, {Miley}, {Morabito}, {O'Sullivan}, \&
  {Prandoni}}]{croston2019}
{Croston}, J.~H., {Hardcastle}, M.~J., {Mingo}, B., {et~al.} 2019, \aap, 622,
  A10

\bibitem[{{Fanaroff} \& {Riley}(1974)}]{fanaroff74}
{Fanaroff}, B.~L. \& {Riley}, J.~M. 1974, \mnras, 167, 31P

\bibitem[{{Garon} {et~al.}(2019){Garon}, {Rudnick}, {Wong}, {Jones}, {Kim},
  {Andernach}, {Shabala}, {Kapi{\'n}ska}, {Norris}, {de Gasperin}, {Tate}, \&
  {Tang}}]{garon2019}
{Garon}, A.~F., {Rudnick}, L., {Wong}, O.~I., {et~al.} 2019, \aj, 157, 126

\bibitem[{{Giacintucci} \& {Venturi}(2009)}]{giacintucci2009}
{Giacintucci}, S. \& {Venturi}, T. 2009, \aap, 505, 55

\bibitem[{{Golden-Marx} {et~al.}(2021){Golden-Marx}, {Blanton},
  {Paterno-Mahler}, {Brodwin}, {Ashby}, {Moravec}, {Shen}, {Lemaux}, {Lubin},
  {Gal}, \& {Tomczak}}]{golden2021}
{Golden-Marx}, E., {Blanton}, E.~L., {Paterno-Mahler}, R., {et~al.} 2021, \apj,
  907, 65

\bibitem[{{G{\'o}mez} {et~al.}(1997){G{\'o}mez}, {Pinkney}, {Burns}, {Wang},
  {Owen}, \& {Voges}}]{gomez1997}
{G{\'o}mez}, P.~L., {Pinkney}, J., {Burns}, J.~O., {et~al.} 1997, \apj, 474,
  580

\bibitem[{{Hill} \& {Lilly}(1991)}]{hill91}
{Hill}, G.~J. \& {Lilly}, S.~J. 1991, \apj, 367, 1

\bibitem[{{Huchra} \& {Geller}(1982)}]{huchra82}
{Huchra}, J.~P. \& {Geller}, M.~J. 1982, \apj, 257, 423

\bibitem[{{Longair}(1971)}]{longair1971}
{Longair}, M.~S. 1971, Reports on Progress in Physics, 34, 1125

\bibitem[{{Massaglia} {et~al.}(2019){Massaglia}, {Bodo}, {Rossi}, {Capetti}, \&
  {Mignone}}]{massaglia2019}
{Massaglia}, S., {Bodo}, G., {Rossi}, P., {Capetti}, S., \& {Mignone}, A. 2019,
  \aap, 621, A132

\bibitem[{{Massaro} {et~al.}(2019){Massaro}, {{\'A}lvarez-Crespo}, {Capetti},
  {Baldi}, {Pillitteri}, {Campana}, \& {Paggi}}]{massaro2019}
{Massaro}, F., {{\'A}lvarez-Crespo}, N., {Capetti}, A., {et~al.} 2019, \apjs,
  240, 20

\bibitem[{{Miley}(1980)}]{miley1980}
{Miley}, G. 1980, \araa, 18, 165

\bibitem[{{Missaglia} {et~al.}(2019){Missaglia}, {Massaro}, {Capetti},
  {Paolillo}, {Kraft}, {Baldi}, \& {Paggi}}]{missaglia2019}
{Missaglia}, V., {Massaro}, F., {Capetti}, A., {et~al.} 2019, \aap, 626, A8

\bibitem[{{Morris} {et~al.}(2022){Morris}, {Wilcots}, {Hooper}, \&
  {Heinz}}]{morris2022}
{Morris}, M.~E., {Wilcots}, E., {Hooper}, E., \& {Heinz}, S. 2022, \aj, 163,
  280

\bibitem[{{O'Donoghue} {et~al.}(1990){O'Donoghue}, {Owen}, \&
  {Eilek}}]{1990ApJS...72...75O}
{O'Donoghue}, A.~A., {Owen}, F.~N., \& {Eilek}, J.~A. 1990, \apjs, 72, 75

\bibitem[{{Owen} \& {Rudnick}(1976)}]{owen1976}
{Owen}, F.~N. \& {Rudnick}, L. 1976, \apjl, 205, L1

\bibitem[{{Prestage} \& {Peacock}(1988)}]{prestage88}
{Prestage}, R.~M. \& {Peacock}, J.~A. 1988, \mnras, 230, 131

\bibitem[{{Smol{\v{c}}i{\'c}} {et~al.}(2007){Smol{\v{c}}i{\'c}}, {Schinnerer},
  {Finoguenov}, {Sakelliou}, {Carilli}, {Botzler}, {Brusa}, {Scoville},
  {Ajiki}, {Capak}, {Guzzo}, {Hasinger}, {Impey}, {Jahnke}, {Kartaltepe},
  {McCracken}, {Mobasher}, {Murayama}, {Sasaki}, {Shioya}, {Taniguchi}, \&
  {Trump}}]{smolcic2007}
{Smol{\v{c}}i{\'c}}, V., {Schinnerer}, E., {Finoguenov}, A., {et~al.} 2007,
  \apjs, 172, 295

\bibitem[{{Tago} {et~al.}(2010){Tago}, {Saar}, {Tempel}, {Einasto}, {Einasto},
  {Nurmi}, \& {Hein{\"a}m{\"a}ki}}]{tago10}
{Tago}, E., {Saar}, E., {Tempel}, E., {et~al.} 2010, \aap, 514, A102

\bibitem[{{Taylor}(2005)}]{taylor05}
{Taylor}, M.~B. 2005, in Astronomical Society of the Pacific Conference Series,
  Vol. 347, Astronomical Data Analysis Software and Systems XIV, ed.
  P.~{Shopbell}, M.~{Britton}, \& R.~{Ebert}, 29

\bibitem[{{Tempel} {et~al.}(2012){Tempel}, {Tago}, \&
  {Liivam{\"a}gi}}]{tempel2012}
{Tempel}, E., {Tago}, E., \& {Liivam{\"a}gi}, L.~J. 2012, \aap, 540, A106

\bibitem[{{Vardoulaki} {et~al.}(2019){Vardoulaki}, {Jim{\'e}nez Andrade},
  {Karim}, {Novak}, {Leslie}, {Tisani{\'c}}, {Smol{\v{c}}i{\'c}}, {Schinnerer},
  {Sargent}, {Bondi}, {Zamorani}, {Magnelli}, {Bertoldi}, {Herrera Ruiz},
  {Mooley}, {Delhaize}, {Myers}, {Marchesi}, {Koekemoer}, {Gozaliasl},
  {Finoguenov}, {Middleberg}, \& {Ciliegi}}]{vardoulaki2019}
{Vardoulaki}, E., {Jim{\'e}nez Andrade}, E.~F., {Karim}, A., {et~al.} 2019,
  \aap, 627, A142

\bibitem[{{Wing} \& {Blanton}(2011)}]{wing2011}
{Wing}, J.~D. \& {Blanton}, E.~L. 2011, \aj, 141, 88

\bibitem[{{Worrall} \& {Birkinshaw}(2000)}]{worrall2000}
{Worrall}, D.~M. \& {Birkinshaw}, M. 2000, \apj, 530, 719

\bibitem[{{York} {et~al.}(2000){York}, {Adelman}, {Anderson}, {Anderson},
  {Annis}, {Bahcall}, {Bakken}, {Barkhouser}, {Bastian}, {Berman}, {Boroski},
  {Bracker}, {Briegel}, {Briggs}, {Brinkmann}, {Brunner}, {Burles}, {Carey},
  {Carr}, {Castander}, {Chen}, {Colestock}, {Connolly}, {Crocker}, {Csabai},
  {Czarapata}, {Davis}, {Doi}, {Dombeck}, {Eisenstein}, {Ellman}, {Elms},
  {Evans}, {Fan}, {Federwitz}, {Fiscelli}, {Friedman}, {Frieman}, {Fukugita},
  {Gillespie}, {Gunn}, {Gurbani}, {de Haas}, {Haldeman}, {Harris}, {Hayes},
  {Heckman}, {Hennessy}, {Hindsley}, {Holm}, {Holmgren}, {Huang}, {Hull},
  {Husby}, {Ichikawa}, {Ichikawa}, {Ivezi{\'c}}, {Kent}, {Kim}, {Kinney},
  {Klaene}, {Kleinman}, {Kleinman}, {Knapp}, {Korienek}, {Kron}, {Kunszt},
  {Lamb}, {Lee}, {Leger}, {Limmongkol}, {Lindenmeyer}, {Long}, {Loomis},
  {Loveday}, {Lucinio}, {Lupton}, {MacKinnon}, {Mannery}, {Mantsch}, {Margon},
  {McGehee}, {McKay}, {Meiksin}, {Merelli}, {Monet}, {Munn}, {Narayanan},
  {Nash}, {Neilsen}, {Neswold}, {Newberg}, {Nichol}, {Nicinski}, {Nonino},
  {Okada}, {Okamura}, {Ostriker}, {Owen}, {Pauls}, {Peoples}, {Peterson},
  {Petravick}, {Pier}, {Pope}, {Pordes}, {Prosapio}, {Rechenmacher}, {Quinn},
  {Richards}, {Richmond}, {Rivetta}, {Rockosi}, {Ruthmansdorfer}, {Sandford},
  {Schlegel}, {Schneider}, {Sekiguchi}, {Sergey}, {Shimasaku}, {Siegmund},
  {Smee}, {Smith}, {Snedden}, {Stone}, {Stoughton}, {Strauss}, {Stubbs},
  {SubbaRao}, {Szalay}, {Szapudi}, {Szokoly}, {Thakar}, {Tremonti}, {Tucker},
  {Uomoto}, {Vanden Berk}, {Vogeley}, {Waddell}, {Wang}, {Watanabe},
  {Weinberg}, {Yanny}, {Yasuda}, \& {SDSS Collaboration}}]{york2000}
{York}, D.~G., {Adelman}, J., {Anderson}, John~E., J., {et~al.} 2000, \aj, 120,
  1579

\bibitem[{{Zirbel}(1997)}]{zirbel1997}
{Zirbel}, E.~L. 1997, \apj, 476, 489

\end{thebibliography}

\end{document}